# Can High-Temperature Reactions Be Described by a Minimum Energy Path Model? Steric Hindrance Matters

Xiongzhi Zeng[1†], Zongyang Qiu[1†], Pai Li[1†], Zhenyu Li[1]* and Jinlong Yang[1]

[1] Hefei National Laboratory for Physical Sciences at the Microscale, University of Science and Technology of China, Hefei 230026, P. R. China.

* Corresponding author: zyli@ustc.edu.cn
† X.Z., Z.Q., and P.L. contributed equally to this work.

## Abstract

High-temperature reactions widely exist in nature. However, they are difficult to be characterized either experimentally or computationally. The routinely used minimum energy path (MEP) model in computational modeling of chemical reactions is not justified to describe high-temperature reactions since high-energy structures are actively involved there. In this study, using $CH_4$ decomposition on the Cu(111) surface as an example, we systematically compare MEP results with those obtained by explicitly sampling all relevant structures via *ab initio* molecular dynamics (AIMD) simulations at different temperatures. Interestingly, we find that, for reactions protected by a strong steric hindrance effect, the MEP is still effectively followed even at a temperature close to the Cu melting point. In contrast, without such a protection, the flexibility of surface Cu atoms can lead to a significant free energy barrier reduction at a high temperature. Accordingly, some conclusions about graphene growth mechanisms based on MEP calculations should be revisited. Physical insights provided by this study can deepen our understanding on high-temperature surface reactions.

## Introduction

Precisely characterizing the elementary chemical reactions is a prerequisite to understand many complex processes, such as heterogeneous catalysis and nanostructure growth. According to the transition state theory, [1] reaction rate is exponentially dependent on the barrier height defined as the free energy difference between the transition state and the reactant state. As a common practice, this free energy barrier is estimated by identifying the minimum energy path (MEP) on the potential energy surface (PES). MEP starts from the lowest-energy structure of the reactant state and passes through a saddle point on the PES which is the lowest-energy structure of the transition state. Usually, the activation energy which is the energy difference between these two structures makes a dominant contribution to the free energy barrier. Other temperature-dependent contributions can then be estimated with a harmonic-oscillator or ideal-gas approximation. [2,3] Such a MEP based protocol is computationally very effective and it is thus routinely used in reaction studies even when the reaction temperature is high. However, the MEP model is questionable at a high temperature since structures far away from the MEP become also important.

Describing high-temperature reactions correctly is especially important in methane and hydrogen fed graphene chemical vapor deposition (CVD) growth on Cu surfaces, [4–6] since the typical growth temperature (1000 °C) is approaching the melting point of Cu (1084.62 °C). Due to the complexity of the growth process, most previous theoretical studies on graphene growth still use the MEP model to describe elementary chemical reactions. [7–13] MEP calculations predict that methane decomposition on Cu surface is thermodynamically and kinetically very difficult, especially for the last dehydrogenation step (CH → C + H). [14] Accordingly, methane decomposition is considered to be the rate-determining step of graphene growth and CH is identified as the dominant feeding species of graphene growth when the $H_2$ partial pressure is high. [15] These important conclusions about graphene growth become questionable if MEP based thermodynamics and kinetics are not reliable. Therefore, it is highly desirable to go beyond the MEP model[16] and systematically study methane decomposition on Cu surfaces.

In this study, methane decomposition on the Cu(111) surface is investigated by comparing results from MEP calculations and from ab initio molecular dynamics (AIMD) simulations at different temperatures. Since it explicitly samples all relevant configurations, AIMD can be applied to both low- and high temperature reactions. For $CH_4$ dehydrogenation, AIMD simulations indicate that gas-phase reactions are involved. Therefore, it should not be well described by an MEP model without a gas-phase contribution and we will focus more on the three surface reactions, $CH_i$ ($i$ = 1-3) dehydrogenation. At 300 K, AIMD free energy barriers for these three reactions can be well reproduced by adding temperature-dependent corrections to MEP energy barriers, which confirms that MEP is a reasonable model at this temperature. At 1300 K, with a melting Cu surface, one may not expect MEP to be a good approximation anymore. However, the MEP based barriers are surprisingly still very close to the AIMD results for $CH_2$ and $CH_3$ dehydrogenation. Coordination number analysis suggests that, in these two cases, although the overall surface structure is totally different, the local chemical environment around the reaction center is still similar to the low-temperature case due to a steric hindrance effect from H atoms. In contrast, CH dehydrogenation is a reaction not well protected by the steric hindrance effect, where AIMD simulations show that the C atom can be wrapped by up to six Cu atoms at this temperature, significantly different from MEP structures. As a result, the AIMD free energy barrier is much lower than that predicted by MEP for CH decomposition. Kinetic Monte Carlo (KMC) simulations indicate that such a difference can change the mechanism of graphene growth especially for those under high $H_2$ partial pressures. These results provide us useful insights in understanding high temperature surface reactions.

**Computational Details**

All the first-principles calculations were performed using the Vienna ab initio simulation package (VASP) [17] with the Perdew-Burke-Ernzerhof exchange-correlation functional. [18] Projector augment wave method [19] was used with a plane-wave energy cutoff of 500 eV. The Cu(111) surface was modeled by a four-layer 4 × 4 supercell with a 15 Å vacuum layer. The Brillouin zone was sampled with a 3 × 3 × 1 k-grid. Transition states were located using the climbing image nudged elastic band (NEB) method. [20]

Due to the high energy barriers, C-H bond breaking is very difficult to be observed in a brute-force AIMD simulation. Enhanced sampling was realized using the metadynamics [21] technique by adding history-dependent bias potentials in a predefined collective coordinate space. Potential of mean force (PMF) was obtained from a metadynamics simulation using the PLUMED package. [22] Cu coordination number of atom $i$ was defined as a continuous function of the distance to neighboring Cu atom $j$

$$CN = \sum_j \frac{1-\left(\frac{r_{ij}}{r_0}\right)^{18}}{1-\left(\frac{r_{ij}}{r_0}\right)^{36}}$$

The switching parameter $r_0$ was chosen to be 2.4 and 2.0 Å for carbon and hydrogen, respectively. Test calculations indicate that results obtained in this study are not very sensitive to these parameters.

Graphene growth on the Cu surface at 1300 K was simulated using the standard rejection-free KMC approach. [23,24] A 2000×2000 lattice was used to represent the Cu(111) surface. To speed up KMC simulations, a mean field approximation was applied to H adatoms since they have a high concentration and a rapid diffusion rate. Only the number of H adatoms was recorded and the hydrogenation and dehydrogenation rates were adjusted accordingly. More details about the KMC simulations can be found in our previous studies. [15]

**Results and Discussion**

**Dehydrogenation at 300 K**. MEPs of $CH_4$, $CH_3$, $CH_2$, and CH dehydrogenation on Cu (111) are identified via NEB calculations, which gives activation energies of 1.63, 1.55, 1.03, and 1.88 eV, respectively, agreeing well with previous results. [15,25–27] These results suggest that CH decomposition is the most difficult step in the sequential $CH_4$ dehydrogenation reactions. AIMD metadynamics simulations are performed at 300 K using a specific C-H distance as the collective coordinate to obtain the PMF (Supporting Information Figure S1). In AIMD trajectories, the overall structure of the Cu(111) surface is well maintained and Cu atoms oscillate around their equilibrium positions. As expected, the obtained free energy barriers for $CH_i$ (i=1-4) dehydrogenation are different from the MEP activation energies (Figure 1a), since there is no temperature effect included in the latter case.

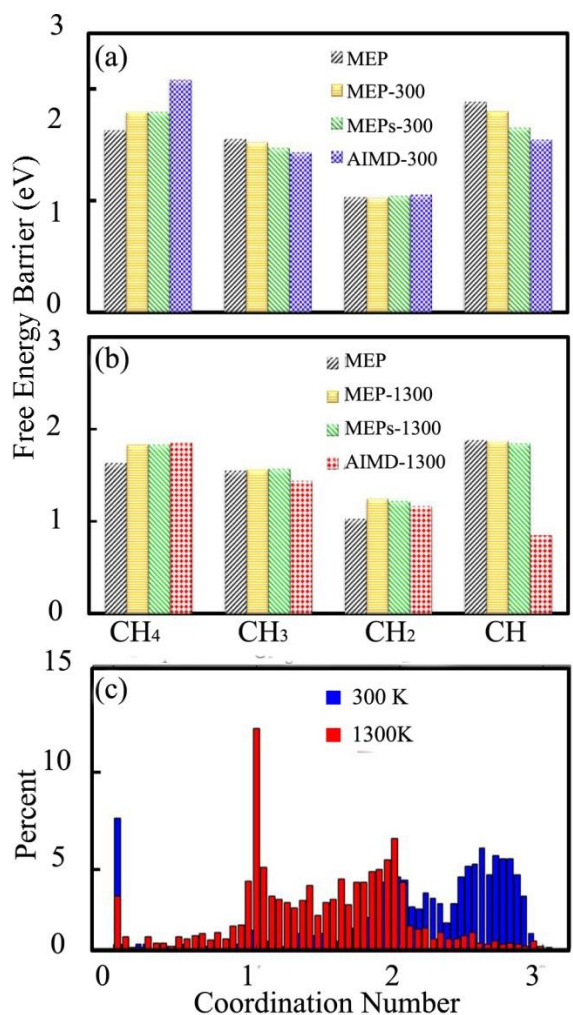

**Figure 1** Activation energy (black) and free energy barrier (yellow, green, red, and blue) in $CH_4$, $CH_3$, $CH_2$, and CH dehydrogenation at (a) 300 and (b) 1300 K. (c) Distribution of the transition-state Cu coordination number of H in $CH_4$ dehydrogenation AIMD trajectories at different temperatures.

To directly compare with AIMD results, on top of the MEP activation energy, some corrections should be applied to obtain an estimation of the free energy barrier at 300 K. We first discuss the dehydrogenation of $CH_3$, $CH_2$, and CH. Vibrational contributions are added within a harmonic approximation to obtain an MEP-based free energy barrier estimation (MEP-300). Since nuclear quantum effects are not included in our AIMD simulations, to make the MEP-300 results consistent with AIMD results, vibrational free energy are calculated with a classical harmonic oscillator model, even though the free energy barriers obtained from classical and quantum harmonic oscillator models are similar (typically within a 0.1 eV difference, Supporting Information Figures S19). For $CH_3$ and $CH_2$ dehydrogenation, the MEP-300 estimation of the free energy barrier is already agreed well with the AIMD results (Figure 1a). However, there is still a small discrepancy in the CH dehydrogenation case.

An MEP typically starts from the most stable structure of the reactant, which corresponds to the fcc hollow-site adsorption for $CH_2$ and CH dehydrogenation and the hcp hollow site adsorption for $CH_3$ dehydrogenation. However, in AIMD simulation, more than one adsorption sites can be visited due to the relatively low diffusion barriers. [28] Therefore, in addition to the MEP associated with the most stable reactant structure, MEPs starting from other relevant structures should also be considered. [29] By simply averaging over all relevant MEP barriers (Supporting Information Figure S15) using the Boltzmann weight factor of their initial state structures, we obtain an effective free energy barrier (MEPs-300). MEPs-300 results for $CH_3$, $CH_2$, and CH dehydrogenation all agree well with their corresponding AIMD results and the maximum free energy barrier difference is smaller than 0.11 eV. Therefore, the MEP model can describe surface reactions correctly when the temperature is not very high, which is the reason why it is nowadays routinely used in kinetics studies.

In the $CH_4$ dehydrogenation case, $CH_4$ in the reactant state is mainly in the gas phase during the AIMD simulation (Supporting Information Figure S10). Therefore, in principle, it is not possible to compare the AIMD result with the MEP model since gas phase states have not been incorporated into the latter case. Nevertheless, for completeness, we still make an MEP-300 free-energy barrier correction to the MEP activation energy. One thing different here is that the translational and rotational degrees of freedom of the reactant-state $CH_4$ are treated with the ideal gas model, [2,3] since the diffusion and rotation barriers of $CH_4$ on Cu(111) are extremely small. The resulting MEP-based estimation of $CH_4$ dehydrogenation barrier is 0.29 eV lower than the AIMD result. Such a discrepancy mainly comes from the gas phase dehydrogenation contribution in the AIMD results. As shown in Figure 1c, there is a notable probability that $CH_4$ in the transition state is still in the gas phase (Cu coordination number of H atoms is almost zero). Considering that the gas phase dehydrogenation of $CH_4$ has an energy barrier as high as 4.12 eV (Supporting Information Figure S13), even the possibility of gas phase dehydrogenation is as low as 7.6% (Supporting Information Figure S12), a significant increase of the free energy barrier (around (4.12-1.63)×7.6% as a very rough estimation) is still expected. Notice that the possibility of gas phase dehydrogenation is expected to have been overestimated in our AIMD simulation [30] due to an underestimation of the van der Waals interaction in the density functional theory adopted here.

**Dehydrogenation at 1300 K.** When we increase the temperature to a value (1300 K) close to what was adopted in graphene growth, the ordered Cu(111) structure at 300 K becomes completely destroyed and the structures observed in AIMD trajectories (Figure 2) are generally very different from those in MEPs. As will be discussed in more details below, in the CH dehydrogenation case, fluctuation of the Cu coordination number is significantly larger than other dehydrogenation steps. To speed up the convergence, we use the number of bridging Cu atoms between C and H as an additional collective coordinate in the metadynamics simulation of CH dehydrogenation. The resulting two-dimensional free energy surface can then be projected back to the C-H distance dimension. Figure 1b gives all free energy barriers obtained from AIMD simulations at 1300 K.

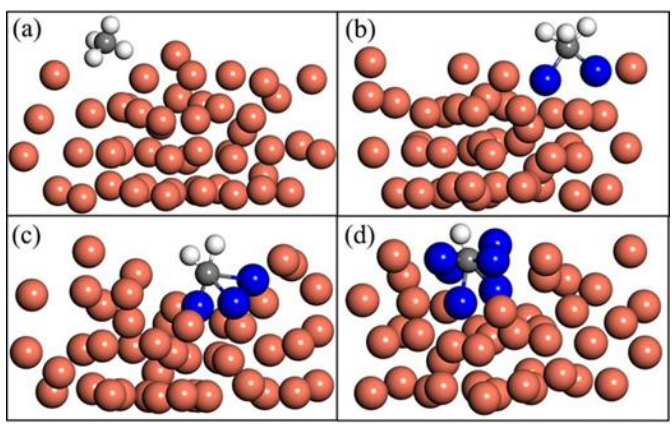

**Figure 2** A typical reactant-state snapshot of the AIMD trajectory for (a) $CH_4$, (b) $CH_3$, (c) $CH_2$, and (d) CH dehydrogenation at 1300 K.

Since AIMD structures are now very different from the ordered Cu(111) based MEP structures, it is interesting to see how big a difference will be there between an AIMD free energy barrier and a free energy barrier estimated from MEP results. Surprisingly, the MEP-1300 barriers still agree well with the AIMD results for $CH_3$ and $CH_2$ dehydrogenation (Figure 1b) even though AIMD structures are very different. This can be understood by checking the local chemical environment around the reaction center. Although the Cu surface is highly disordered at 1300 K, the local structure of adsorbed $CH_2$ or $CH_3$ is still similar to that in the ordered Cu(111) case. For example, as shown in Figure 2c, $CH_2$ in that snapshot has a structure similar to the hollow site adsorption on a smooth Cu(111) surface.

The situation in the CH dehydrogenation case, however, is different. It turns out that CH can be easily trapped in a cavity created by the fluctuation of Cu atoms on the surface, which makes its local chemical environment significantly different from that on a flat crystalline surface. The difference between the $CH_3$/$CH_2$ and CH cases is mainly a steric hindrance effect. Although Cu atoms on the surface are quite flexible, due to the existence of multiple H atoms, there is only a very limit space for Cu atoms to approach the C atom in $CH_3$ or $CH_2$. In contrast, CH is not well protected by the steric hindrance effect, which makes it be easily trapped in a cavity formed by Cu atoms. Since CH can coordinate with more Cu atoms on a melting surface, the dehydrogenation free energy barrier is significantly lowered compared to that in the low-temperature case. It also makes the CH dehydrogenation free energy barriers predicted by MEP and AIMD very different at 1300 K, since large surface distortion is not included in the MEP model.

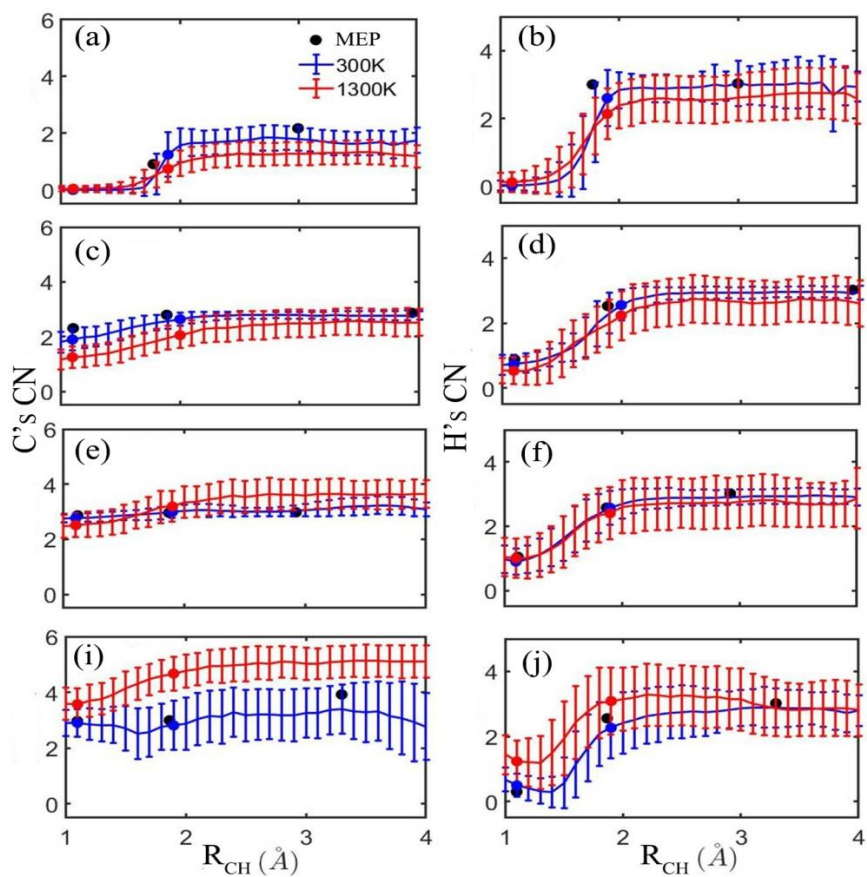

**Figure 3** Average Cu coordination numbers of C and H atoms in (a-b) $CH_4$, (c-d) $CH_3$, (e-f) $CH_2$, and (i-j) CH dehydrogenation reactions as a function of the C-H distance at different temperatures. The initial, transition, and final states are marked by circles.

To more systematically investigate the local chemical environment, we perform a Cu coordination number analysis for each AIMD trajectory (Figure 3). Notice that some widely used concepts referring to a specific surface structure, such as active site, are more difficult to be applied to a dynamic surface system. Cu coordination number depends on both the number of neighboring Cu atoms and the distances to them. For $CH_i$ ($i$ = 1-3) statically adsorbed at a hollow site on Cu(111), where the number of neighboring Cu atoms is fixed to three, the latter plays a determining role. Due to the difference of Cu-C distances, the Cu coordination number of C varies from 2.31 for $CH_3$ to 2.86 for $CH_2$ and it reaches 2.96 for CH. On a disordered surface at 1300 K, the coordination number of C is mainly determined by the number of neighboring Cu atoms. In fact, its mean value (about 1.25 for $CH_3$, 2.41 for $CH_2$ and 3.47 for CH) roughly follows the number of dangling bonds in $CH_i$ ($i$ = 1-3). For $CH_4$ molecule which is weakly adsorbed on the substrate, the coordinate number is almost zero at all temperatures.

Generally, the Cu coordination numbers of both C and H increases along the reaction pathways with an elongation of the C-H bond, due to a stronger interaction with the surface at a longer C-H distance. How the Cu coordination number of C changes with the temperature depends on the steric hindrance effect. For species with a strong steric hindrance, such as $CH_3$, only a limited space is available for Cu atoms to approach C. To effectively utilize this space, ordering of neighboring Cu atoms is essential. Therefore, flat surface on which stable adsorption is observed as in the MEP case can be more effectively coordinated with the adsorbate compared to melting surface. As a result, Cu coordination number decreases at elevated temperature (Figure 3c). The steric hindrance in $CH_2$ is relatively weaker compared to $CH_3$ and the Cu coordination numbers at 300 and 1300 K are more or less the same.

For the smallest species, CH, the Cu coordination number of C at 1300 K is significantly larger than that at 300 K or zero-temperature. The difference at the transition state is especially large, which indicates a strongly different local chemical environment. Such a difference is also reflected in the Cu coordination number of H, where a much more significant temperature effect is observed in the CH decomposition case compared to the cases of $CH_2$ and $CH_3$ dehydrogenation. It is the main reason why the free energy barrier predicted by AIMD at 1300 K is very different from that estimated from MEP. In fact, the local chemical environment effect can be observed even when the surface remains to be ordered. For example, if we use the MEP model to study CH decomposition on Cu(111), Cu(100), and Cu(410) (with an increasing coordination number), we can find that a larger coordination number gives a lower barrier (Supporting Information Figure S18). [31]

Gas phase reaction is no longer negligible for $CH_4$ dehydrogenation at 1300 K. [30] However, in AIMD simulations, the possibility of the reactant-state $CH_4$ molecule to be in the gas phase is decreased comparing to that in the 300 K case (Supporting Information Figure S12). Notice that a wall potential is applied on top of the surface to avoid $CH_4$ being diffused to be far away from the surface. Therefore, geometry fluctuation of surface Cu atoms will increase the chance of $CH_4$ to be contacted with Cu atoms. Again, for $CH_4$ dehydrogenation, a comparison of AIMD results including gas-phase reactions (Figure 1c) with MEP results without a gas-phase contribution is in principle impossible. However, the estimation of $CH_4$ dehydrogenation free energy barrier (MEPs-1300) obtained using the same protocol as in the 300 K case surprisingly agrees with the AIMD result. This is actually an error cancellation effect. On one hand, in the reactant state, the ideal gas model is not expected to describe well the rotational and translational degrees of freedom of $CH_4$ on a fluctuating surface, which gives an over-stabilized reactant state. On the other hand, since there is no gas phase dehydrogenation contribution, the transition state in the MEP model is also over-stabilized.

Using free energy barriers shown in Figure 1, rate constants of the dehydrogenation reactions can be estimated with the transition state theory. [1,29,32] As listed in Table I, the dehydrogenation rates at 1300 K are always higher than those at 300K. Generally, rates predicted from AIMD and MEP based free energy barriers are similar except for $CH_4$ dehydrogenation at 300 K and CH dissociation at 1300 K. The former is mainly an effect of gas-phase reaction. Here, we focus on pure surface reactions, and then the largest discrepancy comes from the 1300 K CH decomposition, where there is a four orders of magnitude difference between MEP and AIMD results. Because MEP fails to describe the high-temperature local chemical environment when the steric hindrance effect is weak, CH decomposition rate at 1300 K is strongly underestimated by the MEP model.

**Table I.** Reaction rate constant for $CH_i$ ($i$=1-4) dehydrogenation at 300 (I) and 1300 (II) K obtained from the MEP model and PMF in AIMD simulations. All values are in $s^{-1}$.

|     |     | $CH_4$    | $CH_3$   | $CH_2$  | $CH$     |
|-----|-----|-----------|----------|---------|----------|
| I   | MEP | 5.31E-18  | 1.26E-12 | 2.11E-5 | 1.19E-15 |
|     | PMF | 1.27E-21  | 2.71E-11 | 6.99E-5 | 1.16E-14 |
| II  | MEP | 2.17E6    | 2.22E7   | 5.04E8  | 1.99E6   |
|     | PMF | 4.81E6    | 1.37E8   | 2.46E9  | 4.96E10  |

**Graphene growth modeling.** Since the MEP model is routinely used in previous theoretical studies on graphene growth, it is important to check the effect of the underestimation of CH decomposition rate on graphene growth mechanisms. For this purpose, KMC simulations with the CH dehydrogenation rate updated with the PMF result are performed. Other reactions involved in graphene growth are expected to be reasonably well described with the MEP model. For example, attachment/detachment of carbon species at/from graphene island edges is expected to be protected by the steric hindrance effect from the graphene island. Nevertheless, the simple protocol adopted here with only CH dehydrogenation rate updated is not expected to give reliable graphene growth mechanisms. The main purpose of the KMC simulations presented here is to demonstrate how large the effect of more accurate kinetic parameters can be.

KMC simulations are performed at 1300 K under different $H_2$ and $CH_4$ partial pressures using MEP kinetic parameters in case A. In case B, the same parameters are used except that the CH decomposition rate is updated with the AIMD result. As an example, results under 10 Torr of both $H_2$ and $CH_4$ partial pressures are shown in Figure 4. In both cases, sequential dehydrogenation steps converting $CH_4$ to CH are observed. However, the reaction pathways after the formation of CH become very different. Generally, there are three possible subsequent pathways, i.e. further dehydrogenation, CH combination to form $C_2H_2$, and graphene edge CH attachment [15]. Further dehydrogenation to form C monomer is suppressed in case A. However, it dominates in case B. Although CH attachment to graphene edge occurs 419 times in case B, there is also 412 times of CH detachment observed in the KMC trajectory. Therefore, the net CH attachment (7 times) is negligible comparing to CH dehydrogenation. For CH combination to form $C_2H_2$, we even observe that its reverse reaction occurs 2 more times. The dominance of the CH decomposition pathway can be easily understood from the much fast CH decomposition in case B compared to that in case A.

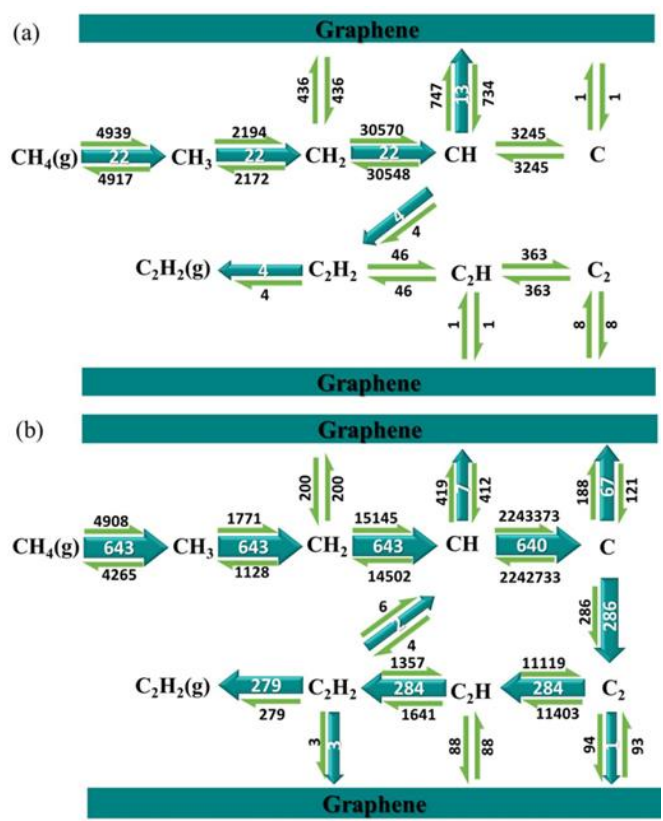

**Figure 4** (a) Kinetic pathways from KMC simulations in case A using kinetic parameters from MEP energy barriers. Both $H_2$ and $CH_4$ pressures are 10 Torr. Diffusion on the surface and $H_2$ adsorption/desorption are not shown. Gas phase species are marked with a 'g' character. Black numbers are occurring times of reactions and reverse reactions indicated by the green arrows, while white numbers give the net occurring times. (b) KMC kinetic pathways in case B with the CH decomposition rate obtained from PMF.

From the growth mechanism point of view, a fundamental issue is to identify the dominant feeding species of graphene growth. Previously, on the basis of MEP based kinetic parameters, $C_2$ or CH have been identified as the dominant feeding species of graphene growth under low and high $H_2$ partial pressures. [15,33] This conclusion should be revisited, since CH decomposition has been strongly underestimated previously. Faster CH decomposition leads to a decrease of CH steady-state concentration. As shown in Figure 4b, even when the $H_2$ partial pressure is already high, CH may still not be the dominant feeding species of graphene growth.

Notice that results presented here do not mean all previous studies on graphene growth are suspicious. Thanks to the physical insight obtained here about the effect of steric hindrance, we can distinguish conclusions which are likely to be still valid from those should be revisited. For example, the hydrogen saturation induced stabilization of graphene edge is mainly determined by rates of attachment/detachment reactions at metal passivated and hydrogen saturated graphene edges. [15] These reactions are protected by a strong steric hindrance from the graphene island. Therefore, such conclusions, although previously obtained based on the MEP model, are expected to be reliable.

**Conclusion**

In summary, we have performed AIMD simulations to study $CH_4$ dissociation on Cu surface, focusing on how well the MEP model can be used to describe high-temperature reactions. It turns out that it is still a reasonable approximation even when the surface is already melting as long as there is a strong steric hindrance protection. For small-species reactions not well protected by steric hindrance, such as CH dehydrogenation, the free energy barrier predicted by AIMD simulation can be significantly lower than the value predicted by MEP based models. Therefore, some conclusions drawn previously on graphene growth based on overestimated CH decomposition barrier should be revisited. This study provides useful insights in understanding high temperature reactions.

**Acknowledgements**

This work was partially supported by NSFC (21825302), MOST (2016YFA0200604), and by USTC-SCC, Tianjin, and Guangzhou Supercomputer Centers.